\documentclass[]{aa}  
\usepackage{graphicx}
\usepackage[dvipsnames]{xcolor}
\usepackage[colorlinks=true,breaklinks=true]{hyperref}
\usepackage{natbib}
\usepackage{amsmath}
\let\oldAA\AA
\renewcommand{\AA}{\text{\normalfont\oldAA}}

\definecolor{lightblue}{rgb}{0.11,0.66,1}

\hypersetup{linkcolor=lightblue,citecolor=lightblue,filecolor=cyan,urlcolor=cyan}

\begin{document}

   \title{Introducing the LBT Imaging of Galactic Halos and Tidal Structures (LIGHTS) survey} 
   
   \titlerunning{First results of the LIGHTS survey}
\authorrunning{Ignacio Trujillo et al.}

   \subtitle{A preview of the low surface brightness Universe to be unveiled by LSST}

   \author{Ignacio Trujillo\inst{1,2}, Mauro D'Onofrio\inst{3}, Dennis Zaritsky\inst{4}, Alberto Madrigal-Aguado\inst{1,2}, Nushkia Chamba\inst{5}, Giulia Golini\inst{3},
             Mohammad Akhlaghi\inst{1,2}, Zahra Sharbaf\inst{6}, Ra\'ul Infante-Sainz\inst{1,2}, Javier Rom\'an\inst{1,2,7},  Carlos Morales-Socorro\inst{8}, David J. Sand\inst{4}, Garreth Martin\inst{4,9}
           }
   \institute{Instituto de Astrof\'{\i}sica de Canarias,c/ V\'{\i}a L\'actea s/n, E38205 - La Laguna, Tenerife, Spain\\
              \email{trujillo@iac.es}
         \and
            Departamento de Astrof\'isica, Universidad de La Laguna, E-38205 - La Laguna, Tenerife, Spain  
            \and
            Department of Physics and Astronomy, University of Padova, Vicolo Osservatorio 3, I-35122, Italy
            \and
            Steward Observatory, University of Arizona, 933 North Cherry Avenue, Tucson, AZ 85721-0065, USA
            \and
            The Oskar Klein Centre, Department of Astronomy, Stockholm University, AlbaNova, SE-10691 Stockholm, Sweden
            \and
            School of Astronomy, Institute for Research in Fundamental Sciences (IPM), P.O. Box 1956836613, Tehran, Iran
            \and
            Instituto de Astrofísica de Andalucía (CSIC), Glorieta de la Astronomía, E18008, Granada, Spain
            \and
            Universidad Internacional de Valencia, C/ Pintor Sorolla, 21, E46002, Valencia, Spain
            \and
            Korea Astronomy and Space Science Institute, 776 Daedeokdae-ro, Yuseong-gu, Daejeon 34055, Korea
             }

   \date{Received September 15, 1996; accepted March 16, 1997}

 
  \abstract
   {We present the first results of the LBT Imaging of Galaxy Haloes and Tidal Structures (LIGHTS) survey. LIGHTS is an ongoing observational campaign with the 2$\times$8.4m Large Binocular Telescope (LBT) aiming to explore the stellar haloes and the low surface brightness population of satellites  down to a depth of  $\mu_V$$\sim$31 mag/arcsec$^2$ (3$\sigma$ in 10\arcsec$\times$10\arcsec boxes) of nearby galaxies.  We  simultaneously collected deep imaging in the \textsl{g} and \textsl{r} Sloan filters using the Large Binocular Cameras (LBCs). The resulting images are 60 times (i.e. $\sim$4.5 mag) deeper than those from the Sloan Digital Sky Survey (SDSS), and they have characteristics  comparable (in depth and spatial resolution) to the ones expected from the future Legacy Survey of Space and Time (LSST). Here we show the first results of our pilot programme targeting NGC1042 (an M33 analogue at a distance of 13.5 Mpc) and its surroundings. The depth of the images allowed us to detect an asymmetric stellar halo in the outskirts of this galaxy whose mass (1.4$\pm$0.4$\times$10$^8$ M$_{\odot}$) is in agreement with the  $\Lambda$ Cold Dark Matter ($\Lambda$CDM) expectations. Additionally, we show that deep imaging from the LBT  reveals  low mass satellites  (a few times 10$^5$  M$_{\odot}$) with very faint central surface brightness $\mu_V$(0)$\sim$27 mag/arcsec$^2$ (i.e. similar to Local Group dwarf spheroidals, such as Andromeda XIV or Sextans, but at distances well beyond the local volume). The depth and spatial resolution provided by the LIGHTS survey open up a unique opportunity to explore the `missing satellites' problem in a large variety of galaxies beyond our Local Group down to masses where the difference between the theory and observation (if any) should be significant.}

   \keywords{galaxies: evolution -- galaxies: formation -- galaxies: halos -- galaxies: photometry -- galaxies: structure -- Cosmology: dark matter}

   \maketitle
%

\section{Introduction}

The imminent arrival of new large-aperture, wide-field of view telescopes covering a significant fraction of the night sky is set to revolutionise many facets of present-day astronomy. Deep imaging is certainly no exception. The possibility of deep imaging of large areas of the sky will allow us to make statistical studies in many astrophysical branches. While this becomes a reality, there are a number of astrophysical problems requiring deep imaging with high spatial resolution where current large telescopes can provide significant insights ten years in advance. With this goal in mind, we present the first results of the LBT Imaging of Galaxy Haloes and Tidal Structures (LIGHTS) survey. The main objectives of the LIGHTS survey  are to address two problems closely related to the nature of dark matter: the properties of  stellar haloes of galaxies and the population of faint satellites around them. The lessons to be learned from LIGHTS will pave the way for the massive statistical analysis we expect from the Vera C. Rubin observatory by the end of the Legacy Survey of Space and Time (LSST) survey.

The presence of a low surface brightness stellar halo around  galaxies is an unavoidable consequence of the galaxy formation mechanism within the $\Lambda$ Cold Dark Matter ($\Lambda$CDM) cosmological paradigm. Dark matter haloes are built through multiple mergers \cite[][]{1978MNRAS.183..341W}, and during this process, they get disrupted together with their stellar component. This results in a low surface brightness stellar halo associated with the surviving galaxy, a record of the accretion history of the object
\citep[][]{2005ApJ...635..931B,2006MNRAS.365..747A,2008ApJ...689..936J}. The star counting technique (either with space- or ground-based telescopes) has been used to explore the stellar properties of haloes belonging to the closest galaxies to a high level of detail  \citep[see e.g.][]{2005ApJ...631..262R,2007AJ....134...43H,2009MNRAS.395..126I,2009Natur.461...66M,2010ApJ...718.1118D,2011ApJ...738..150T,2013ApJ...766..106M,2014ApJ...780..128I,2015ApJ...809L...1O,2016ApJ...823...19C,2017MNRAS.466.1491H,2019ApJ...886..109C,2020ApJ...905...60S}. However, this technique remains mostly limited to galaxies where individual stars can be resolved \citep[i.e. up to 16 Mpc using current technology; see e.g.][]{2012MNRAS.421..190Z}. Therefore, if we want to significantly increase  the number of explored galaxies, we need to rely on deep broad-band imaging of unresolved stellar populations. Although there are marvellous examples of bright streams surrounding galaxies \citep[e.g.][]{2010AJ....140..962M,2018A&A...614A.143M}, a comprehensive analysis of the stellar haloes  of galaxies (showing its intricate structure) requires images deeper than $\mu_V$=30 mag/arcsec$^2$  \citep[see e.g.][]{2008ApJ...689..936J,2010MNRAS.406..744C}. This depth is around a factor of 2.5 to ten times (i.e. 1 to 2.5 magnitudes) deeper  than current deep broad-band imaging \citep[see e.g.][]{2012ApJS..200....4F,2013ApJ...762...82M,2015MNRAS.446..120D,2015A&A...581A..10C,2016MNRAS.456.1359F,2016ApJ...830...62M,2018MNRAS.475.3348H,2018ApJ...857..144H,2019MNRAS.490.1539R}. In fact, there is only a handful of galaxies observed at such depth (i.e. $\mu_V$$\gtrsim$30 mag/arcsec$^2$;  3$\sigma$ in 10\arcsec$\times$10\arcsec) with broad-band imaging \citep[e.g.][]{2010A&A...513A..78J,2016ApJ...823..123T}. This surface brightness regime is, however, well within the reach of large telescopes using a modest amount of time (i.e. a few hours of integration time).  Although the techniques and the telescopes are available for such an enterprise,  a systematic analysis of nearby galaxies using wide-field and large aperture telescopes is currently absent. In this sense, our knowledge of the properties of galactic stellar haloes is still in its infancy.

Together with the stellar halo, the $\Lambda$CDM paradigm predicts a large number of dark matter subhaloes coexisting with the host dark matter halo \citep[see e.g.][]{2008MNRAS.391.1685S,2008Natur.454..735D}. For about twenty years, however, it has been known that the number of observed satellites around galaxies is significantly smaller than the number of predicted dark matter subhaloes  \citep[][]{1999ApJ...524L..19M,1999ApJ...522...82K}. The discrepancy gets significantly worse at the low-mass end. However, this `missing satellites' problem\footnote{When the baryon physics is taken into account then the missing satellites problem becomes much less severe \citep[see e.g.][]{2018A&A...616A..96R}.} is a result of exploring the abundance of dwarf satellites around our own Galaxy \citep[see e.g.][]{2012AJ....144....4M} and Andromeda  \citep[see e.g.][]{2016ApJ...833..167M,2018ApJ...868...55M}. To investigate how representative this problem is in a cosmological context, it is absolutely necessary to explore the satellite population around other massive galaxies beyond the Local Group \citep[see e.g.][]{2021MNRAS.502.1205R} and with different morphologies \citep[see e.g.][]{2015MNRAS.454.1605R}. In recent years, there has been an intense effort in this direction \citep{2021arXiv210501658M}, with studies of the satellite population of galaxies down to M$_V$$\sim$-8.5 mag in galaxies like Centaurus A \citep[see e.g.][D=3.6 Mpc]{2019A&A...629A..18M,2019ApJ...872...80C}, M94 \citep[][D=4.2 Mpc]{2018ApJ...863..152S} and M101 \citep[see e.g.][D=7 Mpc]{2019ApJ...885..153B,2019ApJ...878L..16C,2020ApJ...893L...9B}. \citet{2020ApJ...891..144C,2021ApJ...908..109C} have extended this analysis up to a distance of D=12 Mpc and completeness down to M$_g$$\sim$-10 mag. Finally, the SAGA survey is exploring the satellite population (down to M$_r$$\sim$-12.3 mag) in Milky Way analogues up to a redshift of z$\sim$0.01 \citep{2021ApJ...907...85M}.

In this work, we  show that ultra-deep imaging from LBT  is capable of detecting satellites down to M$_V$=-8.25 mag and with central surface brightness of $\mu_g(0)$=27.1 mag/arcsec$^2$ at a distance of 13.5 Mpc. By targeting fainter systems, the contrast of not just faint, but low surface brightness
objects should increase relative to the background galaxy population. This potential improvement on the statistical determination of which fraction of faint galaxies are real satellites promises to significantly extend our capabilities to explore the `missing satellites' problem  well beyond the Local Volume (i.e. at distances beyond 12 Mpc).

The LIGHTS survey aims to systematically detect stellar haloes and `missing satellites' galaxies (if any)  of a representative sample of nearby  (D$\sim$20 Mpc) galaxies  by obtaining ultra-deep imaging (i.e. $\mu_V$$\gtrsim$30 mag/arcsec$^2$;  3$\sigma$ in 10\arcsec$\times$10\arcsec). These observations are performed using the LBC cameras of the LBT telescope in the Sloan bands \textsl{g} and \textsl{r}. In this paper, we present the first results of one galaxy in the LIGHTS sample, NGC1042.

This paper is structured as follows: in Section 2, we describe the properties of  NGC1042 and its surroundings. We present the data used in Section 3. Section 4 is dedicated to the analysis of the results.  Finally we discuss and summarise our results in Section 5. All the magnitudes are provided in the AB system.

 \section{NGC1042 and its surroundings}

NGC1042 is a Sc  galaxy \citep{1981RSA...C...0000S} located at R.A.(2000)=02h40m24.0s and Dec(2000)=-08d26m01s. The galaxy distance $D=13.5\pm2.6$ Mpc has been recently measured using the Tully-Fisher relationship \citep[][]{2019ApJ...880L..11M}. That distance agrees very nicely with the one estimated by \citet[13.2 Mpc;][]{2007A&A...465...71T} using the Large Scale Reconstruction of the region. At such a distance, 1\arcsec\ corresponds to 0.064 kpc.

The galaxy is located in a region of the sky with relatively low dust contamination \citep[A$_g$=0.095 and A$_r$=0.065 mag; ][according to NASA/IPAC Extragalactic Database, NED]{1998ApJ...500..525S,2011ApJ...737..103S}. Using SDSS, \citet{2019ApJ...880L..11M} measured the following apparent magnitudes for NGC1042: 11.24$\pm$0.05 (\textsl{g}-band) and 10.80$\pm$0.05 (\textsl{r}-band), which after correction for Galactic extinction implies a global colour of \textsl{g}-\textsl{r}=0.41 mag. The absolute magnitude in the \textsl{r}-band is -19.9 mag.
Using \citet{roediger2015}, we estimate a (M/L)$_\textsl{r}$ for the galaxy of $\sim$0.75$\Upsilon_{\odot}$. Consequently, the stellar mass is estimated to be $\sim$5$\times$10$^9$ M$_{\odot}$. This is around 1/10th of the stellar mass of the Milky Way \citep[see e.g.][]{2015ApJ...806...96L}.

The  H{\sc i} mass can be obtained from its integrated, self-absorption-corrected, H{\sc i} line flux: S$_{H{\sc I}}$=58.12$\pm$5.90 Jy km s$^{-1}$ \citep[][]{2005ApJS..160..149S}.
We apply the following equation to get the H{\sc i} mass M$_{H{\sc I}}$=2.36$\times$10$^5$$\times$D$^2$$\times$S$_{H{\sc I}}$=2.5$\pm$0.4$\times$10$^9$ M$_{\odot}$ \citep[see e.g.][]{filho2013}.
The dynamical mass within the region dominated by the baryonic disc can be roughly measured by using the size of the galaxy and its rotational velocity \citep[M$_{dyn}$(M$_{\odot}$)=2.326$\times$10$^5$R$_{25}$v$_{rot}$$^2$; see e.g.][]{pohlen2006}. The radial distance corresponding to the 25 mag/arcsec$^2$ (\textsl{B}-band) isophote is R$_{25}$=1.95 arcmin \citep[HyperLeda;][]{makarov2014}, or 7.5 kpc. Its maximum rotational velocity corrected for inclination is 128 km/s \citep[see Fig. 14 in ][]{1986AJ.....91..791V} where we have used an inclination of 34.4 degrees \citep[][]{2019ApJ...880L..11M}. Consequently, the dynamical mass of NGC1042 in its innermost $\sim$8 kpc  is M$_{dyn}$=2.9$\times$10$^{10}$ M$_{\odot}$.
Therefore, the galaxy has properties (stellar mass, dynamical mass, etc) similar to M33 \citep[see e.g.][]{corbelli2000}.

The environment around NGC1042 has  recently attracted a lot of attention \citep[see e.g.][]{Muller2019} due to the  claim that this region contains two galaxies `lacking dark matter'  \citep[(KKS2000)04 also known as NGC1052-DF2 and NGC1052-DF4;][]{vD_df2,vD2019}. Unfortunately, neither of these interesting galaxies is in the field of view we are analysing here. A detailed analysis of this area shows that there are two groups of galaxies in projection \citep[][]{2019ApJ...880L..11M,2020ApJ...904..114M}. One is centred around NGC1052 and NGC1047 (both visible in the field of view of  LBT that we study in this work) at a distance of 19 Mpc \cite[][]{2001ApJ...546..681T,2003ApJ...583..712J}, while the other group is at the distance of NGC1042, that is $\sim$13.5 Mpc.  In addition, there are a number of very diffuse galaxies \citep[like NGC1052-DF1;][]{2018ApJ...868...96C} which we  describe in the following sections.

\section{Data}

Ultra-deep observations of the galaxy NGC1042 and its surrounding region were carried out with the Large Binocular Telescope using the Large Binocular Cameras \citep[Blue and Red simultaneously; ][]{2008A&A...482..349G}. The images were taken using Director Discretionary Time (DDT; PI D'Onofrio) during the night of 14th October 2020 in dark conditions. The total amount of time provided was 2h. LBC cameras are composed of 4 CCDs with a pixel scale of 0.2254\arcsec. Each CCD covers approximately 7.8\arcmin$\times$17.6\arcmin, with gaps between the chips of  $\sim$18\arcsec. The final field of view of the LBC cameras is approximately 23\arcmin$\times$25\arcmin. The LBC Blue camera is blue optimised for observations from around 3500 to 6500$\AA$, while the LBC Red camera is red optimised for observations from approximately 5500$\AA$ to 1 $\mu$m. We use the \textsl{g}-SLOAN filter in the LBC Blue and the \textsl{r}-SLOAN filter in the LBC red.  Images were taken under good seeing conditions, producing a final (stacked) image with a full width at half-maximum (FWHM) seeing of $\sim$0.9\arcsec\ in both bands.

\subsection{Observational strategy}

To obtain a background as flat as possible, we follow an observational strategy similar to the one conducted in \citet{2016ApJ...823..123T}. This consists of obtaining a flat-field from the data itself. To do this with enough accuracy, it is necessary to follow a dithering pattern with a step size similar to or larger than the size of the main object under study. Considering that the diameter of NGC1042 is D$_{25}$$\sim$4\arcmin, we  used steps of 5\arcmin.

The total amount of time granted (2h) was split into 
30 different pointings with an exposure time of 180 seconds each. Taking into account the overheads, the total time on-source was 1.5h in each filter. Due to the relatively low amount of time on source,  to maximise the flat-field precision, this time \citep[contrary to the procedure followed in ][]{2016ApJ...823..123T}, we did not rotate the camera but used a single position angle for all  exposures.

\begin{figure*}
   \centering
   \includegraphics[width=\textwidth]{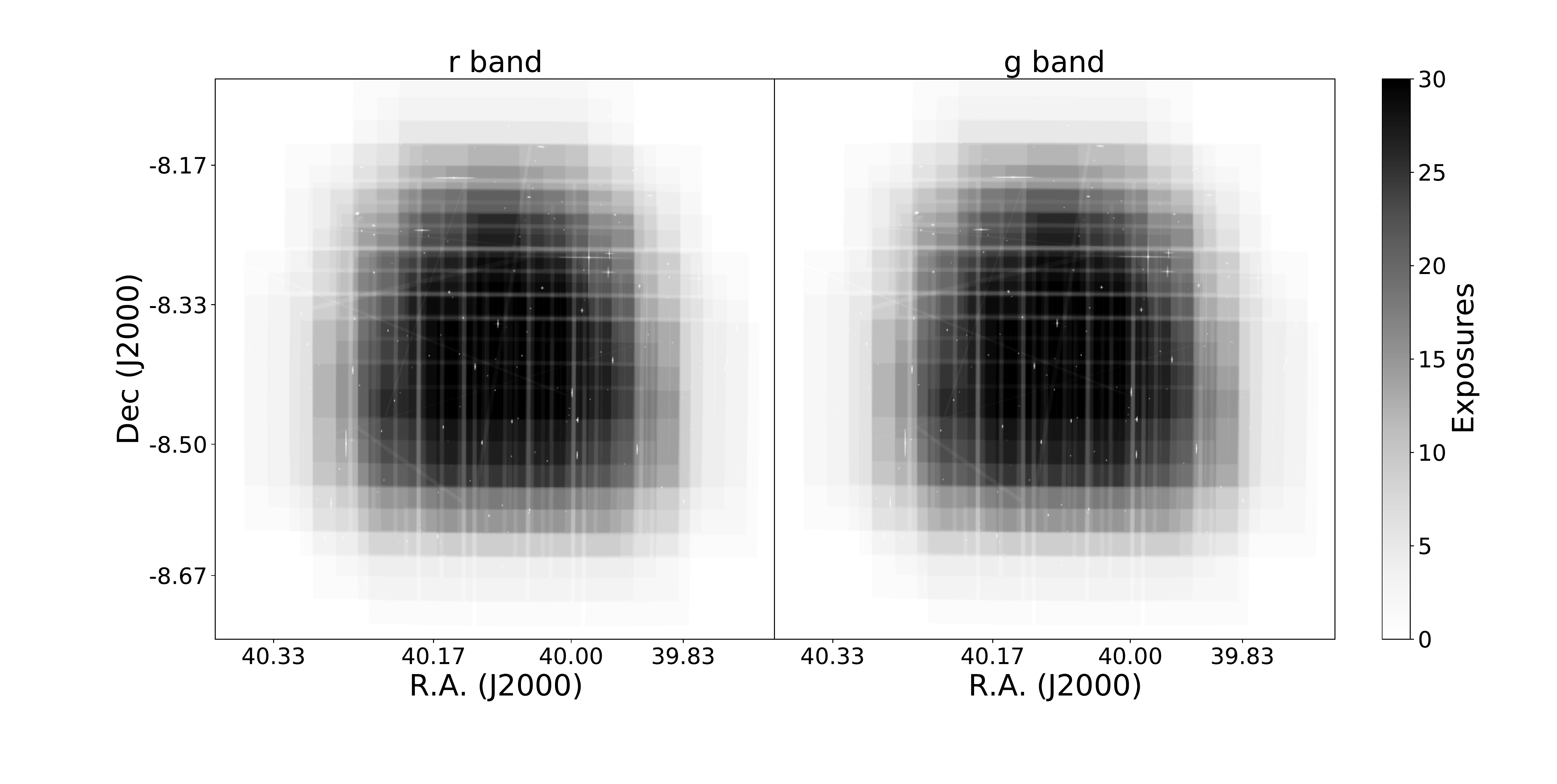}
      \caption{Weight maps in the Sloan \textsl{g}- and \textsl{r}-bands after the stacking process. Darker areas correspond to a larger number of repetitions and consequently deeper imaging. The dithering pattern followed in this observation is clearly visible. North is up and east to the left.}
         \label{Figweightmaps}
   \end{figure*}

The dithering pattern is illustrated in Fig. \ref{Figweightmaps} where we show the weight maps (for \textsl{g} and \textsl{r} bands) resulting from the stacking process of all the individual frames. The observing pattern is such that the galaxy never is located on the same physical area of the CCD across the 30 exposures. The observing pattern also serves to remove  gaps between the LBC CCDs.

\subsection{Flat-field correction}

Using  the bias images, we create a masterbias by combining all the  bias frames with a sigma clipping median using
   \texttt{Gnuastro}\footnote{\url{https://www.gnu.org/software/gnuastro}}'s
   \texttt{Arithmetic}. This is done for each CCD of the LBC. This masterbias is later  subtracted from all  science images. 

Masterflats are created for each filter using the science (already bias corrected) images  themselves. This is performed in two steps. First, the science images are normalised as follows. The pixel (1176,3027) of the central CCD of the LBC (CCD 2) is used as the centre of a normalisation ring. That ring has a inner radius of 2305 pixels and a width of 200 pixels\footnote{The ring radius is chosen to cover, for all CCDs on the focal plane, a region that has a similar illumination. The width of the ring is selected such that the total number of pixels in it ($\sim$3$\times$10$^6$) is large enough to have a statistically robust way of measuring the  median (at the 0.05\% level).}. That ring crosses the four CCDs.  For each CCD and within the corresponding ring region, we calculate the resistant (to the outliers) median value of the pixels. That value is used to normalise the flux of the CCD for each science image. Then, we combine all the normalised and bias subtracted science images using a sigma clipping median stacking. We get four preliminary masterflats this way, one per CCD. Science images are divided by these first step masterflats. This action allows us to better highlight  the sources in the science images. To get an improved masterflat, we consequently mask all the detected sources using \texttt{NoiseChisel} \citep[][]{2015ApJS..220....1A,2019arXiv190911230A}, also part of \texttt{Gnuastro}, and we median combine all the re-normalised and masked images once again to create a final masterflat (one per CCD). Finally, the individual science images of each filter are divided by their corresponding CCD masterflats.

\subsection{Astrometry, sky determination, and image co-addition}

To determine the astrometry of our individual science images we conduct the following steps. We start by calculating an astrometric solution using \texttt{Astrometry.net} \citep[v0.85; ][]{2010AJ....139.1782L}. We use  Gaia eDR3 \citep[][]{2021A&A...649A...1G} as our astrometric reference catalogue. This produces a first astrometric solution that we consequently improve using \texttt{SCAMP} \citep[v.2.10.0; ][]{2006ASPC..351..112B}. \texttt{SCAMP} reads catalogues that we generate using \texttt{SExtractor} \citep[v.2.25.2; ][]{1996A&AS..117..393B} and calculates the distortion coefficient of the images. As each CCD of the LBC has its own distortions, we run this using the entire block of science images of each CCD. After this step, we run \texttt{SWarp} \citep[2.42.5; ][]{2002ASPC..281..228B} on each individual image to put them into a common grid of 9501$\times$9501 pixels.  The image resampling method is LANCZOS3.

Before co-adding all our science images, we subtract the sky of our individual CCD exposures by masking the signal of each image using \texttt{NoiseChisel}. For each CCD, we calculate the median of the Sky image produced by \texttt{NoiseChisel} and we remove that value from the image\footnote{The values of the main \texttt{NoiseChisel} parameters used for this task are: --tilesize=50,50 --minskyfrac=0.9 --meanmedqdiff=0.01 --snthresh=5.2 --detgrowquant=0.7 --detgrowmaxholesize=1000.}. This very conservative way of subtracting the sky, avoiding any polynomial fits, is done to avoid removing any potential large-scale low surface brightness feature in our data. To stack all our images we use a median combination  using \texttt{Gnuastro}'s \texttt{Arithmetic} program \citep[0.13.12-f50c; ][]{2015ApJS..220....1A}. The co-added image is significantly deeper than any individual image, and therefore, a large number of very low surface brightness features emerge from the noise. These features, which include the extended wings on the stars, the stellar haloes around  galaxies, and Galactic cirri, affect the sky determination of our individual science images.
For this reason, it is necessary to mask these regions and repeat the process of the sky determination in the individual exposures.

\subsection{Photometric calibration}

The photometric calibration of our science images is based
on the photometry of SDSS DR12 images \citep{Alam_2015}. Due to the final size of our FOV ($>$30\arcmin$\times$30\arcmin), we use the SDSS tool `mosaics'\footnote{\url{https://dr12.sdss.org/mosaics}}. We create both SDSS \textsl{g} and \textsl{r}-band imaging. The zeropoint of those images is 22.5 mag. We use around 600 (unsaturated) stars in both SDSS and LBT within our FOV to calibrate  the LBT image photometrically. The stars used range between 18 to 22 mag for both filters. We do not find any need to add a colour term between the SDSS filters and Sloan LBT filters. 

The flux of the stars in the SDSS and LBT images that we have used are obtained using Petrosian magnitudes obtained by \texttt{SExtractor}. As the SDSS images are photometrically calibrated,  we matched the photometric catalogue of the stars in SDSS to the LBT catalogue, resulting in the following zeropoints for the LBT images: ZP$_g$=34.527$\pm$0.006$\pm$0.01 and ZP$_r$=34.111$\pm$0.006$\pm$0.01 mag \citep[the first error bar corresponds to the statistical error and the second error bar is the typical photometrical zeropoint calibration error reported by the SDSS team for these filters;][]{2004AN....325..583I}. Therefore, in practice the ultimate photometric precision in the LIGHTS survey is given by the SDSS photometric accuracy. The  final combination is shown in Figure \ref{Figngc1042}. The final images for each filter separately are shown in the Appendix. The limiting surface brightness (3$\sigma$; 10\arcsec$\times$10\arcsec, i.e. equivalent to a 3$\sigma$ fluctuation with respect to the background of the image in an area of 10\arcsec$\times$10\arcsec) of the LBT images are: 31.2 mag/arcsec$^2$ (\textsl{g}-band) and 30.5 mag/arcsec$^2$ (\textsl{r}-band). These values correspond to the central (R$<$11.3\arcmin) region of the images where 24 or more pointings overlap.

  \begin{figure*}
   \centering
   \includegraphics[width=\textwidth]{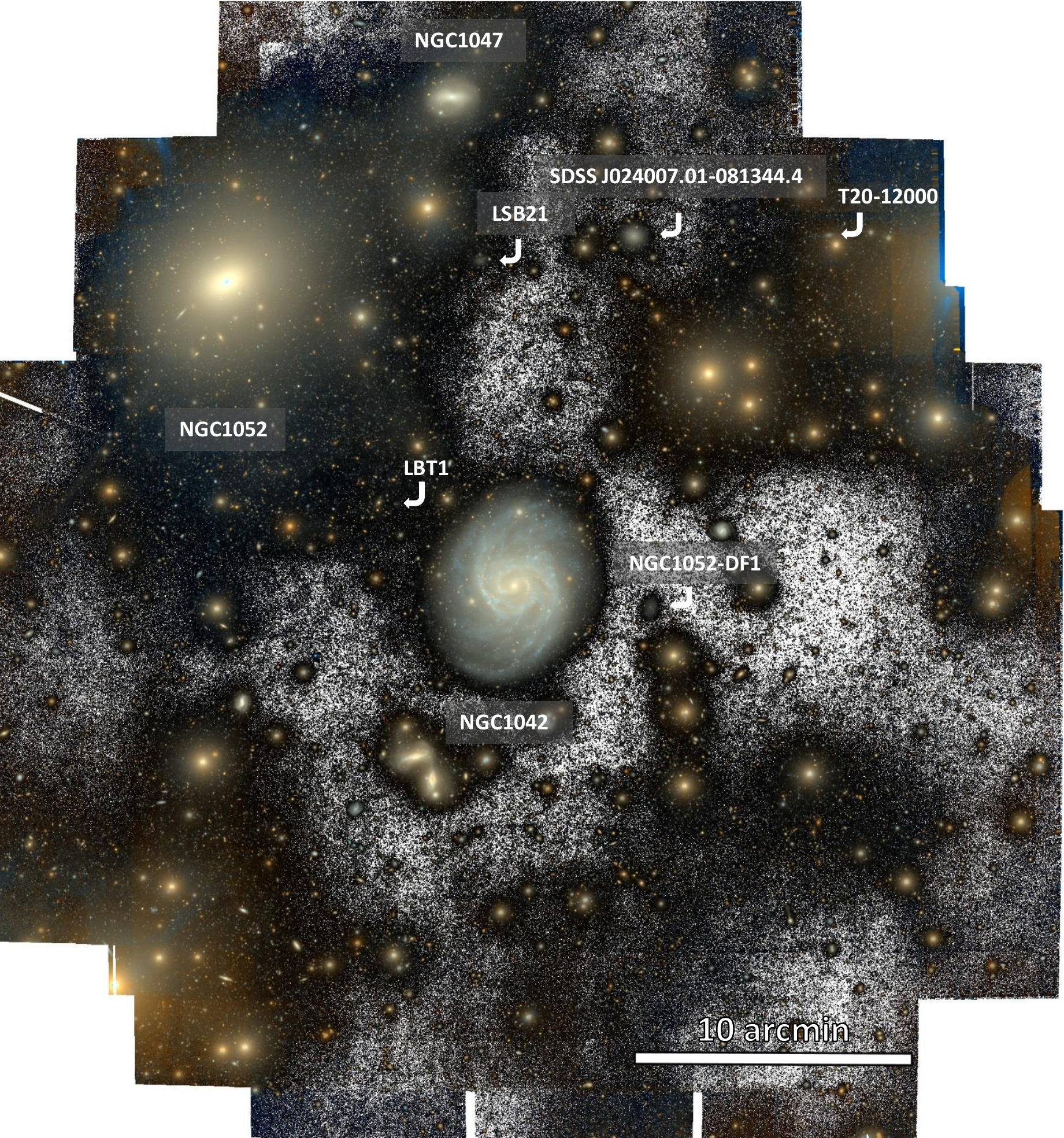}
      \caption{Colour composite image (using \textsl{g} and \textsl{r} Sloan filters) showing NGC1042 and its surroundings. Some known objects and a new identified galaxy (LBT1) are labelled on the figure for reference. The white and black background corresponds to the sum of the \textsl{g} and \textsl{r} filters to enhance the low surface brightness features. The white pixels correspond to the sky region while the darker pixels are the brightest regions. North is up and east to the left.}
         \label{Figngc1042}
   \end{figure*}

\subsection{Other datasets}

Until the  arrival of the next generation of very wide and deep surveys such as the LSST \citep{2019ApJ...873..111I}, SDSS continues to be the reference survey to compare with in terms of imaging due to its excellent sky background quality. On the other hand, images from the Dragonfly telescope array  \citep[][]{2014PASP..126...55A} are commonly cited  as representative of high quality imaging for exploring the low surface brightness Universe. For this reason, to illustrate the depth and quality of our LBT deep imaging, we compare some of the galaxies present in our LBT data with their counterparts on SDSS and Dragonfly imaging\footnote{It is worth noticing the existence of the DESI Legacy Imaging Surveys \citep{2019AJ....157..168D} covering also this region of the sky. \citet{2021arXiv210406071M} has estimated a
limiting surface brightness for this dataset of $\sim$29 mag/arcsec$^2$ (3$\sigma$; 10\arcsec$\times$10\arcsec). Unfortunately, the public available images of this survey are affected by a severe background oversustraction around the brightest galaxies which prevents its use for a direct comparison with our results.}. 

SDSS \textsl{g} and \textsl{r}  band imaging data were retrieved from the DR14 SDSS \citep{2018ApJS..235...42A} Sky Server.
The magnitude zero-point for all the data  is the same: 22.5 mag.
The exposure time of the images is 53.9s and the pixel size 0.396\arcsec.
The observations were taken in drift scanning mode, providing  accurate photometry down to $\mu_r$$\sim$26.5-27 mag/arcsec$^2$ along the surface brightness profiles \citep[\textsl{r}-band;][]{pohlen2006}. Using the following metric (3$\sigma$; 10\arcsec$\times$10\arcsec) for measuring the limiting surface brightness, we achieve  the following SDSS depths for this region of the sky: 27.5 mag/arcsec$^2$ (\textsl{g}-band) and 26.9 mag/arcsec$^2$ (\textsl{r}-band). 

Dragonfly images of the NGC1052 field  \citep[][]{2016ApJ...830...62M} were retrieved from the Dragonfly archive (v0.9)\footnote{\url{https://www.dragonflytelescope.org/data-access.html}}. Images were collected in \textsl{g} and \textsl{r} filters. The number of lenses on the Dragonfly array was 8 at the time the data were taken. Typical exposures for this galaxy were between 15 to 20h. The pixel scale of the  Dragonfly images used here is 2\arcsec. The field of view of the Dragonfly images is $\sim$2$\times$3 degrees. Therefore, we crop these images to cover the same area as that of our LBT dataset. To remove the sky, we masked the Dragonfly images and we subtract a constant value. Later, we calibrate them to the SDSS photometry finding the following zeropoints:  ZP$_g$=16.02 and ZP$_r$=15.86 mag. The limiting surface brightness (3$\sigma$; 10\arcsec$\times$10\arcsec) of the Dragonfly images in the field of NGC1042 are  28.7 mag/arcsec$^2$ (\textsl{g}-band) and 28.0 mag/arcsec$^2$ (\textsl{r}-band)\footnote{\citet[][section 2.2]{2016ApJ...830...62M} measured the limiting surface brightness of these data in boxes of 12\arcsec$\times$12\arcsec\ and with 1$\sigma$ significance  in the \textsl{g}-band. They found a value that ranges from 28.6 to 29.2 mag/arcsec$^2$. Transforming these values to 3$\sigma$ and 10\arcsec$\times$10\arcsec\ corresponds to: 27.2 to 27.8 mag/arcsec$^2$. In other words, their estimation is around 1 mag brighter than our estimation. We think this can be connected to the way they estimate the limiting surface brightness. They use boxes of  12\arcsec$\times$12\arcsec  and measured the variation in the average flux contained therein, while we performed our estimation on a pixel basis, and therefore we are less prone to contamination within those boxes.}. In other words, they are around 1 mag deeper than SDSS images.

 \section{Analysis}

\subsection{The low surface brightness features surrounding NGC1042}

\begin{figure*}
   \centering
   \includegraphics[width=0.75\textwidth]{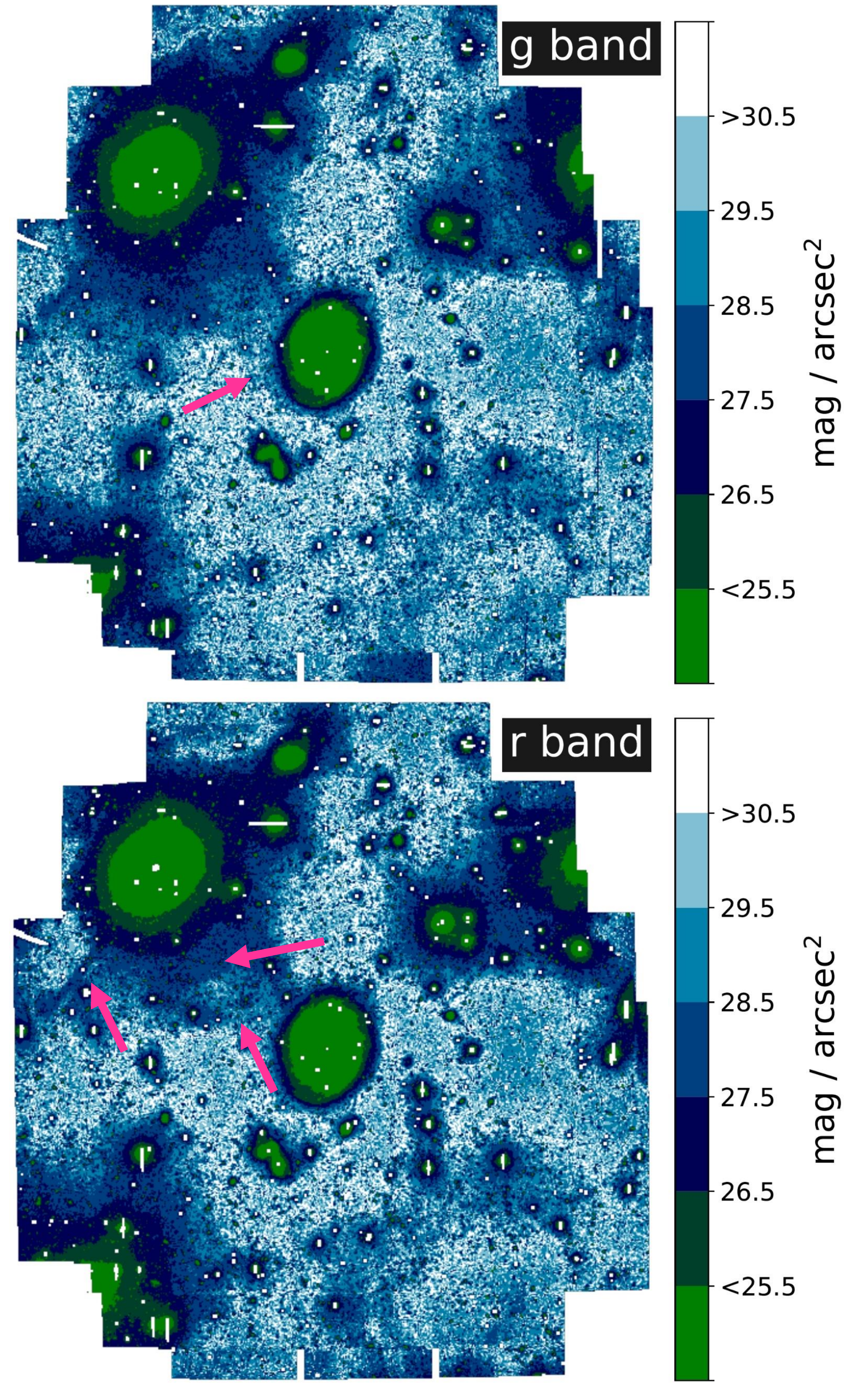}
      \caption{Sloan \textsl{g} (upper panel) and \textsl{r} band (lower panel) images of the NGC1042 region taken with the LBT. The images have been rebinned to a pixel size of 4.5\arcsec (i.e. 20$\times$20 the original pixel scale) to facilitate the observation of the low surface brightness features. The pink arrow on the upper panel points to the extended emission of NGC1042 to the east region. The pink arrows on the bottom pannel indicate the location of stream and shell structures of NGC1052. Towards the edges of these images the number of exposures drop significantly (see Fig. \ref{Figweightmaps}) and the depth and quality decreases.}
         \label{grimages}
   \end{figure*}

Figure \ref{Figngc1042} shows a large number of low surface brightness features surrounding NGC1042. Many of these features were discussed at length in \citet{Muller2019}. For example, one such feature is the stellar bridge between NGC1047 and NGC1052 that indicates their interaction. Another example is the narrow stellar stream on the eastern part of the image, which is likely associated with merging activity in NGC 1052. Of most relevance here is the apparent plume of stars between NGC1042 and NGC1052. Visual inspection at various contrast levels shows that the plume consists of at least two loops of stars belonging to NGC1052 \citep[see also ][]{Muller2019} and so is not a connecting structure between NGC1052 and NGC1042 (Fig. \ref{grimages}). This lack of interaction between these two galaxies is also supported by the absence of a clear distortion in the disc of NGC1042 and excluded given the large measured line-of-sight distance difference: D=13.5 Mpc for NGC1042 \cite[see e.g.][]{2007A&A...465...71T,2019ApJ...880L..11M} and D=19 Mpc for NGC1052 \cite[see e.g.][]{2001ApJ...546..681T,2003ApJ...583..712J}.

\begin{figure*}
   \centering
   \includegraphics[width=1.1\textwidth]{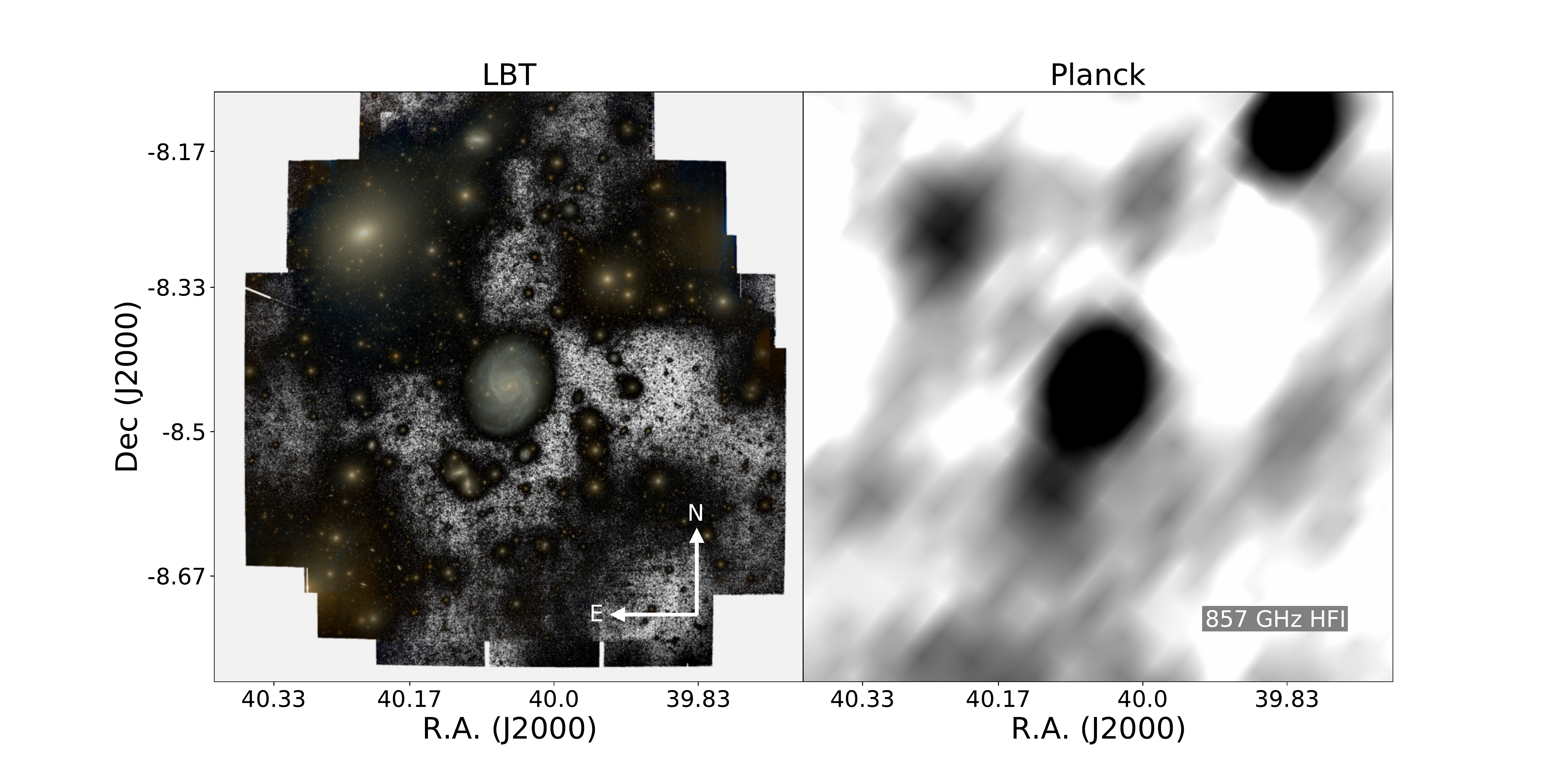}
      \caption{Presence of Galactic cirri in the field of view of NGC1042. The figure compares the optical emission obtained using LBT with the 350 $\mu$m map produced by the Planck satellite \citep{2014A&A...571A...1P}. The location of the dust emission in the Planck channel is qualitatively similar to the position of the lower right extended feature in the optical map. This suggests that the origin of that extended emission could be produced (at least partially) by dust cirri in our own Galaxy.}
         \label{dustplanck}
   \end{figure*}

There is also some light excess towards the bottom right part of the image that is not directly linked to any bright source on the image. For that reason, we have explored whether such brightness excess  can be associated with faint Galactic cirri emission. In Figure \ref{dustplanck}, we show the optical and the radio maps of this region of the sky. We have used the 350 $\mu$m map produced by the Planck satellite to indicate the location of the Galactic cirri \citep{2014A&A...571A...1P}. There is a qualitative  agreement between the location of very extended low surface brightness features in our optical LBT image and the Planck dust map. However, the optical brightness in the bottom right corner is a bit larger than expected according to the dust intensity in this region. For this reason, although we think that part of the excess of light in that corner is real we cannot fully reject the possibility that part of that light results from an artefact of the  data reduction. 

\subsection{A non-symmetric stellar halo around NGC1042}

\begin{figure*}
   \centering
   \includegraphics[width=\textwidth]{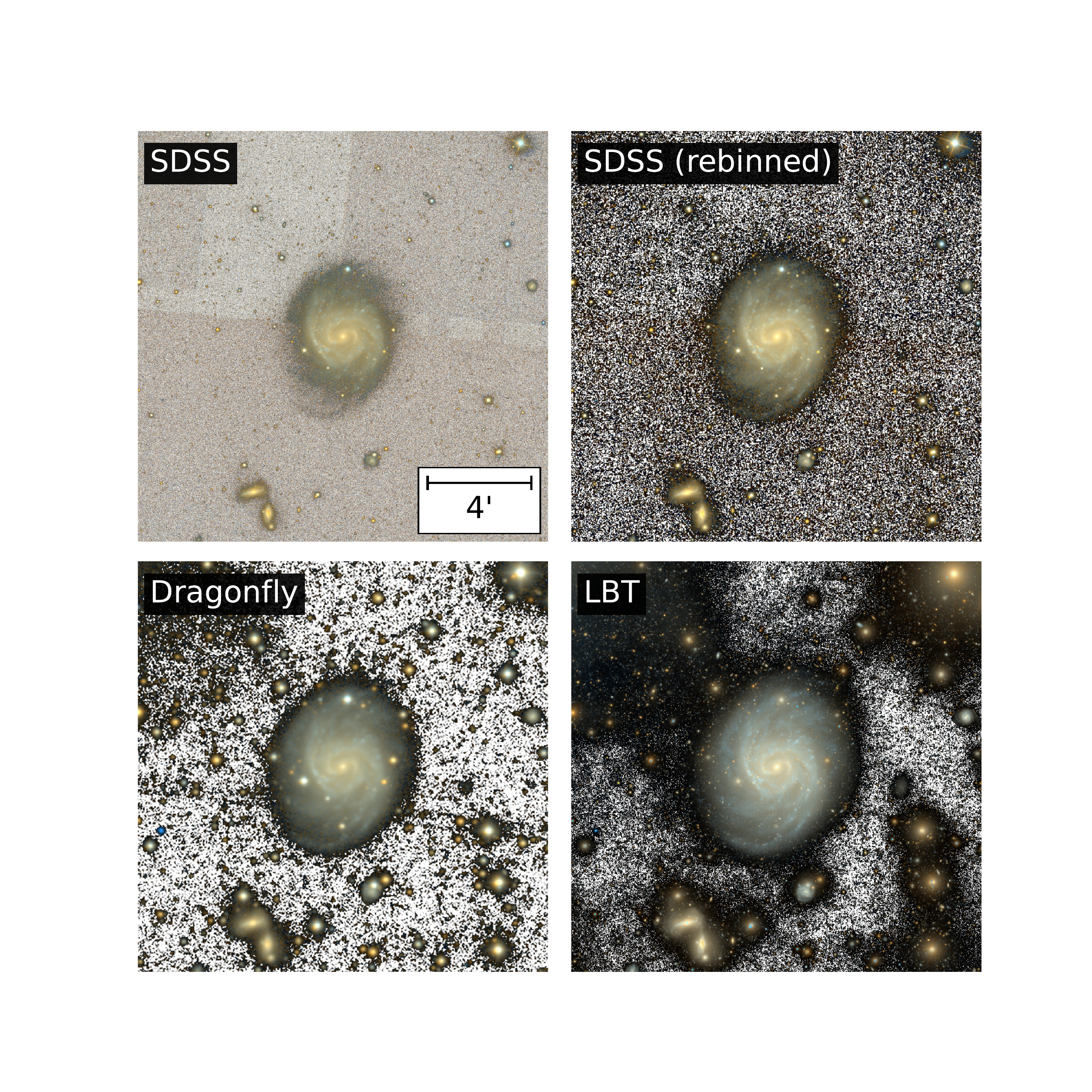}
      \caption{The emergence of the stellar halo of NGC1042 as the depth of  images increases. The panel shows the galaxy NGC1042 at three different surface brightness limiting depths (3$\sigma$; 10\arcsec$\times$10\arcsec): SDSS 26.9 mag/arcsec$^2$ (\textsl{r}-band), Dragonfly 28.0  mag/arcsec$^2$ (\textsl{r}-band) and LBT 30.5 mag/arcsec$^2$ (\textsl{r}-band). In the case of SDSS, we show the visual aspect of the galaxy in two different ways: a) when the original SDSS pixel scale is used (i.e. 0.396\arcsec) and b) when the SDSS data is rebinned to the pixel size of Dragonfly in these images (i.e. 2\arcsec). A non-symmetric stellar halo (with surface brightness $\mu_g$$>$29 mag/arcsec$^2$) is observed in the LBT image. In all the stamps, the background in white and black is built by summing \textsl{g} and \textsl{r} filters to enhance the detection of low surface brightness features. North is up and east to the left.}
         \label{Figngc1042mosaic}
   \end{figure*}

We now focus  on the galaxy NGC1042 itself (Fig. \ref{Figngc1042mosaic}). The object is shown using three different datasets with different depths. As the limiting surface brightness of the data becomes fainter than 30 mag/arcsec$^2$ (3$\sigma$; 10\arcsec$\times$10\arcsec), an excess of light in the outer part of the disc is observed in the eastern region (see Fig. \ref{grimages}). Considering the modest stellar mass of this galaxy ($\sim$5$\times$10$^9$ M$_{\odot}$), similar to M33, the figure shows the importance of reaching very faint limits to observe the stellar haloes of relatively low mass disc galaxies.

\begin{figure*}
   \centering
   \includegraphics[width=\textwidth]{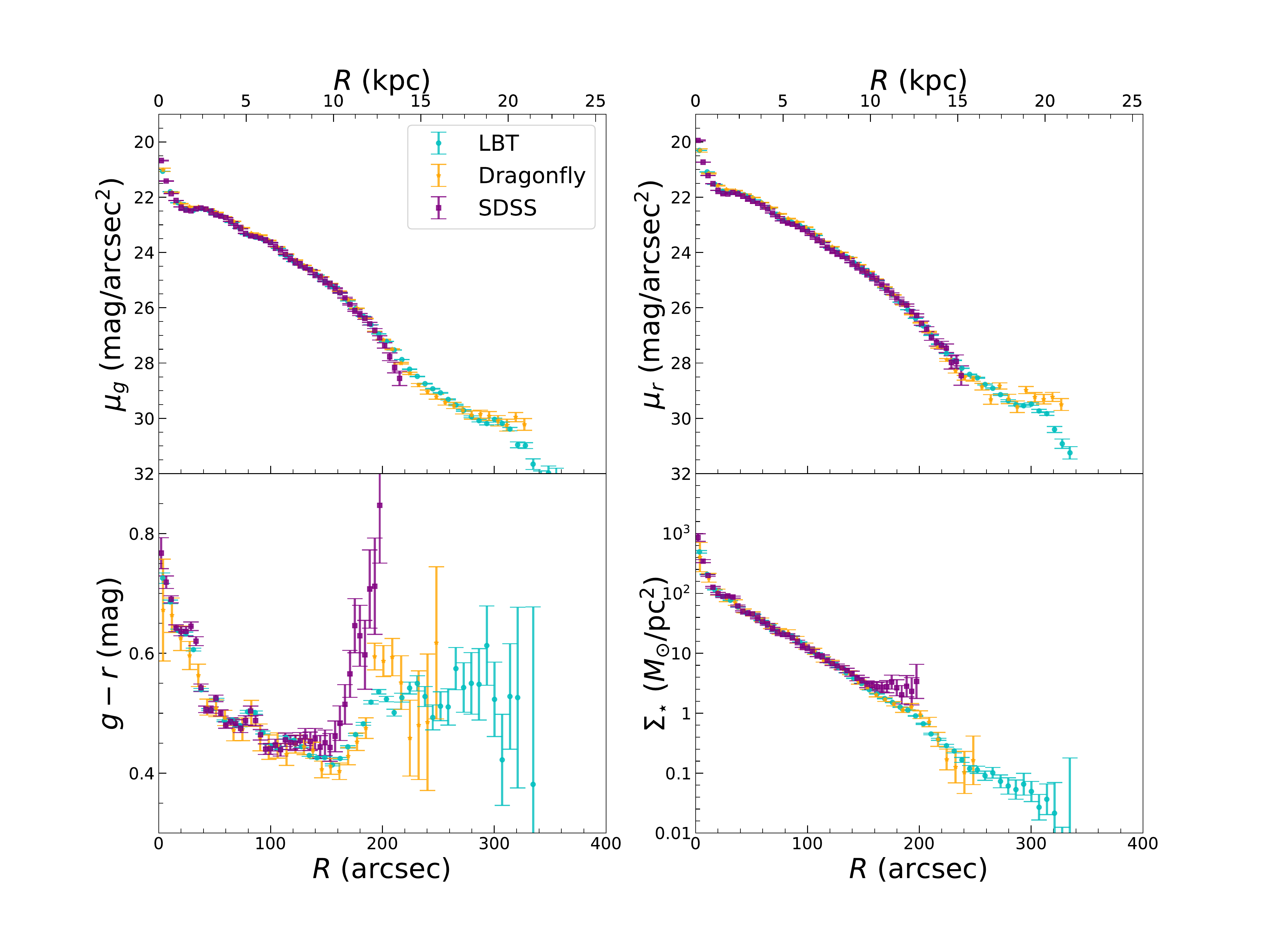}
      \caption{Surface brightness, colour and stellar mass density profiles of NGC1042 using LBT, Dragonfly and SDSS which have different depths. The surface brightness profiles shown here correspond to the observed values while the radial colour and stellar mass density profiles are shown after being corrected by Galactic extinction and the inclination of NGC1042. In the case of the surface brightness profiles, only values above 3$\sigma$ the sky noise of the image are shown. 
      }
         \label{Figprofngc1042}
   \end{figure*}

To quantify the amount of stellar mass in the outer part of NGC1042, we have explored the stellar mass density distribution of the galaxy. To do that, we have first extracted the \textsl{g} and \textsl{r} surface brightness profiles (see Fig. \ref{Figprofngc1042}). This was done after  masking all the surrounding sources using \texttt{NoiseChisel} \citep[we also initially test our masks using the sources detected by \texttt{Max-Tree Objects} finding similar results][]{TeeningaMoschiniTragerWilkinson+2016,2021A&A...645A.107H}. This mask is based on the LBT imaging and it is later also applied to the SDSS and Dragonfly datasets. We  used elliptical apertures with the following properties: axis ratio 0.83 and position angle 15 degrees (measured clockwise from the north axis). These values are the ones provided in \citet{2019ApJ...880L..11M}. In all the data used here, we observe a sudden decrease in the surface brightness profiles of NGC1042 at around 200\arcsec. In the images shown in Fig. \ref{Figngc1042mosaic}, this corresponds to the end of the visible disc in SDSS (rebinned) and Dragonfly datasets. The colour radial profile is  also shown in Fig. \ref{Figprofngc1042}. The shape of the radial colour profile has the characteristic U-shape \citep[see e.g.][]{2008ApJ...684.1026A,2008ApJ...683L.103B,2017MNRAS.467.2127P} of Type II disc galaxies \citep[][]{2005ApJ...626L..81E,pohlen2006}. Interestingly, the end of the U-shape is connected with the sudden drop in the brightness observed in the surface brightness profiles at R$\sim$200\arcsec. We identify this drop as a truncation \citep{1979A&AS...38...15V}. We  expand on the meaning of this truncation feature later on.  Beyond 220\arcsec (i.e. $\sim$14 kpc) the light distribution of NGC1042 no longer follows the symmetry of its disc.  This asymmetric excess of light in the outer part of NGC1042 has a surface brightness of $\sim$29 mag/arcsec$^2$ (\textsl{g}-band). Finally, the LBT surface brightness profiles show a second drop at R$\sim$310\arcsec. For both \textsl{g} and \textsl{r} filters, this radial distance corresponds to a surface brightness of $\sim$30 mag/arcsec$^2$. The reason for this sudden drop in surface brightness in the outer part of the galaxy is connected to the visible end of the asymmetric light around the NGC1042  (see the LBT stamp in Fig. \ref{Figngc1042mosaic}).

Using the surface brightness and radial colour profiles, we calculate the stellar mass to light ratio radial profile using \citet{roediger2015} following the prescription provided in \citet{2008ApJ...683L.103B}.  We use the parameters given in \citet{roediger2015} for a \citet[][BC03]{2003MNRAS.344.1000B}  model and a Chabrier initial mass function \citep[IMF; ][]{2003PASP..115..763C}. The resulting stellar mass density profile of NGC1042 is also shown in Fig. \ref{Figprofngc1042}. A sudden decrease at R=194\arcsec\ is also visible in this profile (which we have identified as a truncation above) which corresponds to a stellar mass density of $\sim$1 M$_{\odot}$/pc$^2$.  This density value is in strong agreement with the suggestion  by \citet{2020MNRAS.493...87T} of using the radial location of this density value as an indicator of  galaxy size. In fact, the truncation  agrees with the idea that this radial location represents the current location of the star formation threshold of this galaxy. Inside this feature we  see a region of active star formation while, beyond this radial position, the stellar light we see  corresponds to the sum of stars that have migrated from the internal disc and/or have been accreted across the history of the galaxy (i.e. its stellar halo). For this last reason, beyond this radial location the light distribution of the galaxy is no longer symmetric. In what follows, we  refer to that radial position as R$_{edge}$\footnote{In a future paper, Chamba et al. (2021, in preparation) we expand on the concept of R$_{edge}$ and its connection with general galaxy properties.}. Consequently, we  measured the amount of stellar mass in the stellar halo as the stellar mass beyond R$_{edge}$.
 
To estimate the total stellar mass of NGC1042, we have integrated its stellar mass density profile down to the radial location where the profile remains reliable (i.e. where the colour profile errors are lower than 0.2 mag). We find the following values: 6.3$\pm$1.6$\times$10$^9$ M$_{\odot}$  (SDSS), 6.2$\pm$1.6$\times$10$^9$ M$_{\odot}$ (Dragonfly) and 6.5$\pm$1.6$\times$10$^9$ M$_{\odot}$ (LBT). This is in good agreement with the raw estimation done in Section 2, using the global galaxy colours provided by \citet{2019ApJ...880L..11M}. Beyond R$_{edge}$, the amount of stellar mass we measure (in the stellar halo) is:  1.2$\pm$0.3$\times$10$^8$ M$_{\odot}$ (Dragonfly) and 1.4$\pm$0.4$\times$10$^8$ M$_{\odot}$ (LBT). Note that for SDSS it is not possible to estimate a stellar halo using our definition as the profile never reaches 1 M$_{\odot}$/pc$^2$ due to its limited depth. Therefore, the fraction of stellar mass in the halo of this galaxy (defined as the fraction of stellar mass beyond R$_{edge}$) is: 2.0$\pm$0.5\% (Dragonfly) and 2.1$\pm$0.5\% (LBT).

To end the discussion on the stellar halo of NGC1042, it is worth noticing that this galaxy was previously identified as a candidate for having a stellar halo (if any) whose stellar mass is below the expectation from the $\Lambda$CDM galaxy formation model \citep[][]{2016ApJ...830...62M,2020MNRAS.495.4570M}. The measurement of the amount of mass in the stellar halo is not a straightforward task. For this reason, in the past, many different approaches have been   conducted to make this measurement (e.g. the mass beyond a given number of effective radii, the mass beyond a given radial physical distance, interpolating the outer shape of the stellar mass density profile towards the inner region, the amount of mass below a given stellar mass density, etc). In \citet{2020MNRAS.495.4570M}, three different ways of estimating the stellar halo mass of NGC1042 were used. In all three cases, the stellar halo mass of the galaxy lies at the bottom edge of the prediction of cosmological simulations (see their Fig. 8). Upon performing a direct comparison with their work, we note the following important difference: the adopted value of the distance to NGC1042. Here we have used a more recent measurement of 13.5 Mpc \citep{2019ApJ...880L..11M}, while \citet{2016ApJ...830...62M} used a previous value of 17.3 Mpc \citep[][]{Tully_2009}. \citet{2016ApJ...830...62M} found a total stellar mass of NGC1042 of 1.53$\pm$0.48$\times$10$^{10}$ M$_{\odot}$ (at 17.3 Mpc). At 13.5 Mpc, that total stellar mass for NGC1042 corresponds to 9.3$\pm$2.9$\times$10$^9$ M$_{\odot}$ (i.e. this value is above the value we measured here using  the Dragonfly data but in agreement within the error bars with our mass estimate for this object).

We can also make a direct comparison of the stellar mass in the halo by measuring the amount of stellar mass  at stellar mass densities below 1 M$_{\odot}$/pc$^2$ \citep[see Fig. 8 in ][]{2020MNRAS.495.4570M}. With this definition, as mentioned above, we find a stellar mass in the stellar halo of 1.2$\pm$0.3$\times$10$^8$ M$_{\odot}$ using the Dragonfly dataset. While \citet{2020MNRAS.495.4570M} find (after the distance correction we have discussed above) 9.6$\pm$2.5$\times$10$^7$ M$_{\odot}$. We find, therefore a 25\% larger value (although still within the error bars). However, when using the LBT dataset, we obtain a stellar halo mass for NGC1042 which is nearly  50\% larger (i.e. 1.4$\pm$0.4$\times$10$^8$ M$_{\odot}$) than the one measured in \citet{2020MNRAS.495.4570M}. The most likely reason for the discrepancy the way the stellar mass density of this galaxy is determined. Both in this work and in \citet{2016ApJ...830...62M}, the stellar mass density is based on the surface brightness profile and the \textsl{g}-\textsl{r} colour radial profile. For that reason, having an accurate colour estimation of the outermost part of the galaxy (in particular in its halo region) is key getting a reliable estimation of the stellar mass in that region. To address this issue, as the Dragonfly S/N is low in that region of the galaxy, \citet{2016ApJ...830...62M} extrapolate outwards the \textsl{g}-\textsl{r} colour they measure in the disc region (R$<$160\arcsec; see their Figure 2). In other words, their outer part of NGC1042 (R$>$160\arcsec) is characterised with a \textsl{g}-\textsl{r} colour that ranges from 0.3 to 0.4 mag. Note that this colour is even bluer than the colour measured in the star forming region of the galaxy. Such a blue \textsl{g}-\textsl{r} colour is in contradiction with the observed colour profile in the Dragonfly dataset (both here and in \citet{2016ApJ...830...62M}) in the region between 160 to 200\arcsec, which shows the well known U-shape up to R=200\arcsec.  Looking now at the LBT colour profile, we see that beyond R=200\arcsec\ the colour of the stellar halo remains red  (\textsl{g}-\textsl{r}$\sim$0.55 mag).  The difference between using a colour of \textsl{g}-\textsl{r}=0.35 mag (i.e. (M/L)$_\textsl{r}$=0.60) instead of \textsl{g}-\textsl{r}=0.55 mag (i.e. (M/L)$_\textsl{r}$=1.27) is a factor of $\sim$2.1 in stellar mass.  Therefore, if \citet{2016ApJ...830...62M} had used a similar colour as the one we measure with the LBT, they would have found a stellar halo mass larger than the value reported. In doing so, the stellar halo mass of NGC1042 will be in better agreement with the expectation from $\Lambda$CDM for this type of galaxy \citep{2020MNRAS.495.4570M}. The present work shows that, to measure the halo stellar masses to better than a factor of two,  colours are essential. Colours can only be obtained well into the halo with observations whose depth allows the exploration of features fainter than 30 mag/arcsec$^2$ (3$\sigma$; 10\arcsec$\times$10\arcsec). To end this section, it is worth mentioning that the effect of the Point Spread Function (PSF) has not  yet been accounted for in the current analysis. Addressing the PSF effect produced by the galaxy itself is key to have an accurate characterisation of the amount of stellar mass in the stellar halo region. To measure the effect of the PSF it is necessary to have a detailed characterisation of the PSF over spatial scales of at least 1.5 times larger than the object under study \citep{2014A&A...567A..97S}.  This will be done in a future work following the prescriptions developed in \cite{2020MNRAS.491.5317I}. Having said that, the low inclination of NGC1042, the fact that both Dragonfly and LBT (with different PSFs) show similar amount of mass (once the colour in the outer part is measured  accurately) in the stellar halo region and the detection of an asymmetric excess of light in the outer part of the galaxy suggest that the effect of the PSF will not be dominant in this particular case.

\subsection{A sample of faint galaxies in the field of view of NGC1042}

\begin{figure*}
   \centering
   \includegraphics[width=\textwidth]{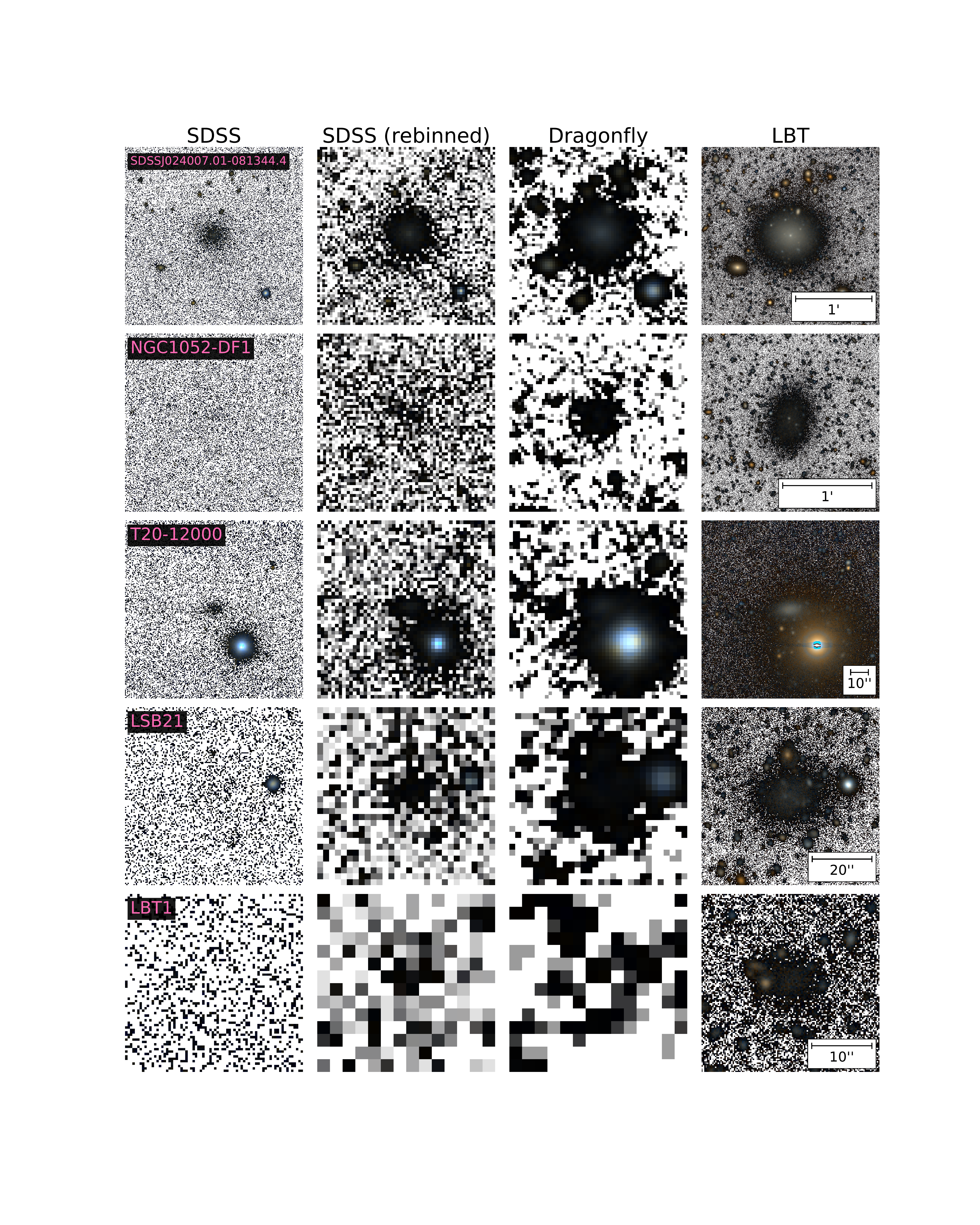}
      \caption{Some examples of low surface brightness galaxies in the field of view of NGC1042. Similar to Fig. \ref{Figngc1042mosaic}, the different stamps show how the galaxies are visible in SDSS (using its native pixel scale 0.396\arcsec), SDSS rebinned to the Dragonfly pixel scale in these images (i.e. 2\arcsec), Dragonfly and LBT.}
         \label{Figsatmosaic}
   \end{figure*}

In the previous section, we have illustrated the extraordinary capacity of LBT to uncover the stellar halo around galaxies, using  NGC1042 as a relatively low mass example. In this section, we show the high potential of the LIGHTS survey to find and explore very low surface brightness (satellite) galaxies. As the aim of this work is only to illustrate the capabilities of LBT for low surface brightness studies, what follows is by no means a comprehensive list of all the low surface brightness objects in this image. In fact, the criteria followed to explore some of these objects are: a) an example of an object being relatively bright but not discussed in the literature as is the case of SDSSJ024007.01-081344.4, b) the faintest low surface brightness galaxy in the field discussed in \citet{2018ApJ...868...96C}:  NGC1052-DF1, c) two low surface brightness galaxies found in deep surveys of the region: T20-12000 \citep{2021ApJS..252...18T} and LSB21 (Rom\'an et al. 2021, in prep), and finally d) a galaxy not previously reported that we have dubbed as LBT1. The summary of the properties of these galaxies are given in Table \ref{table:1} and their are  shown in Figure \ref{Figsatmosaic}.

\begin{figure*}
   \centering
   \includegraphics[width=1.1\textwidth]{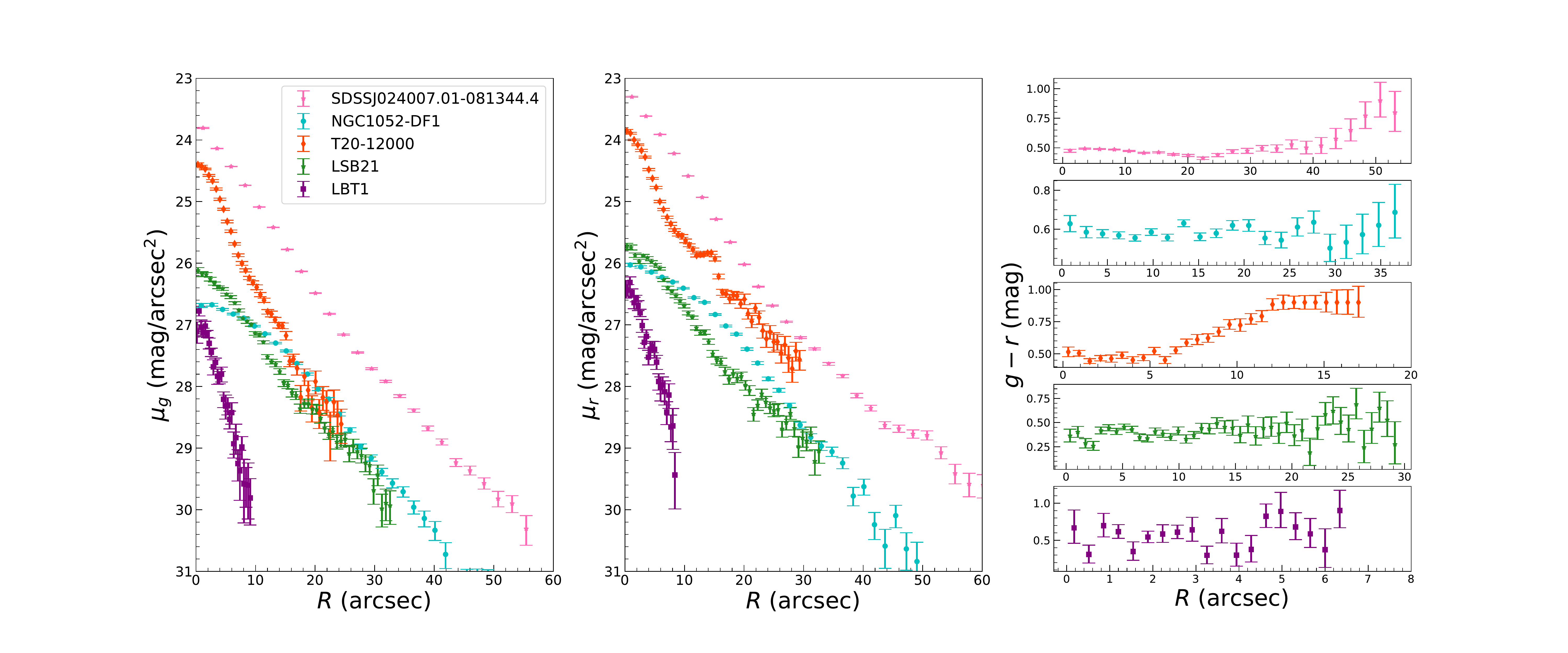}
      \caption{Observed surface brightness profiles of  faint galaxies in the field of view of NGC1042. The left and middle panel shows the observed surface brightness profiles in the \textsl{g} and \textsl{r} bands. The right panel shows the colour radial profiles (corrected for Galactic extinction) down to the distance where the error in their colour values is less than $\pm$0.2 mag.}
         \label{satprofiles}
   \end{figure*}

With the exception of the new object detected here (LBT1) whose size (as measured using the  30 mag/arcsec$^2$ isophote in the \textsl{g}-band) is R$_{30}$$\sim$10\arcsec and whose central surface brightness is $\mu_g$(0)$\sim$27 mag/arcsec$^2$, the rest of the faint galaxies are  visible in the shallower surveys, SDSS and Dragonfly, used here. However, it is only the combination of depth and spatial resolution of LBT that allows us to explore the detailed nature of these galaxies. In fact, for most of these galaxies distinguishing the presence of faint foreground and background objects contaminating the light of these galaxies is only possible with the LBT data. This is key to accurately measuring the structural properties of these objects \citep[see e.g.][]{2017ApJ...850..109B}.
The surface brightness profiles of these faint galaxies and their radial colour profiles obtained with the LBT data are shown in Fig. \ref{satprofiles}. 

\begin{table*}
\caption{Some examples of low surface brightness (satellite) galaxies surrounding NGC1042 explored in this work. The structural properties provided correspond to those found in this work. This table includes the name of the galaxies, their spatial location, their apparent magnitudes in the \textsl{g} and \textsl{r} bands, their central surface brightness, and their effective radii in arcsec.}             
\label{table:1}      
\centering          
\begin{tabular}{c c c c c c c c c c}     
\hline\hline       
Name  & R.A. (2000) & Dec (2000) & m$_g$ & m$_r$ & $\mu_g$(0) & $\mu_r$(0) & R$_{e,g}$ & R$_{e,r}$ & Reference\\ 
& & & (mag) & (mag) & (mag/\arcsec$^2$) &  (mag/\arcsec$^2$) & (\arcsec) & (\arcsec) & \\
\hline                    
      SDSSJ024007.01-081344.4 & 02:40:07.0 & -08:13:44.4 & 17.17 & 16.70 & 23.81 & 23.30 & 11.1 & 10.9 & (1) \\
   NGC1052-DF1 & 02:40:04.6 & -08:26:48.3 & 19.52 & 18.95 & 26.69 & 26.03 & 15.2 & 15.1 & (2)  \\  
   T20-12000 & 02:39:39.3 & -08:13:42.3 & 19.23  & 18.47 & 24.40 & 23.85 & 6.9 & 8.9 & (3) \\
   LSB21 & 02:40:28.8 & -08:14:36.5 & 19.98 & 19.56 & 26.12 & 25.72 & 10.9 & 11.2 & (4) \\
      LBT1 & 02:40:40.6 & -08:23:08.9  & 22.70 & 22.06 & 27.11 & 26.41 & 4.2 & 4.9 & (5) \\
\hline                  
\end{tabular}
\tablefoot{To the best of our knowledge the above galaxies were first found/discussed in the following references: (1) \citet{2003AJ....126.2081A}, (2) \citet{2018ApJ...868...96C}, (3) \citet{2021ApJS..252...18T}; object also known as SDSS J023939.36-081342.0,  (4)  Rom\'an et al. (2021, in prep); object also known as SDSS J024028.61-081436.7, (5) This work. The apparent magnitudes and surface brightness have been corrected by the following Galactic extinction: A$_g$=0.095 and A$_r$=0.065 mag.  The quantities are given showing only the significant figures up to which the values can be regarded as reliable.}
\end{table*}

A brief discussion of the properties of these faint galaxies in the field of NGC1042 follows.

SDSSJ024007.01-081344.4. This is a galaxy with spheroidal appearance and with a central excess of light that is potentially a star cluster. The central surface brightness in the \textsl{g}-band is close to 24 mag/arcsec$^2$. Its \textsl{g}-\textsl{r} colour is very homogeneous and around 0.5 mag. Only in its very outer region there is a red trend visible, which is likely contamination by other nearby sources.
 
NGC1052-DF1. This galaxy has a very low central surface brightness $\mu_g$(0)$\sim$26.7 mag/arcsec$^2$. This object is visible in the SDSS image after re-binning it to the Dragonfly pixel scale. However, its very low surface brightness prevented it from being catalogued in the SDSS archive. The  depth of LBT imaging allow us to trace the structure of this faint  galaxy to more than 3 times its effective radius. The colour of NGC1052-DF1 is very homogeneous \textsl{g}-\textsl{r}$\sim$0.55 mag. There are no signatures of tidal distortions down to the explored radial range (R=50\arcsec) and surface brightness of $\sim$31 mag/arcsec$^2$.
  
T20-12000. This galaxy has an elongated spheroidal appearance. Its vicinity to a brighter star makes it  difficult to explore its outer part. In fact, we observe a large difference in the shape of the surface brightness profiles between the \textsl{g} and \textsl{r} bands. Therefore, the profiles remain unreliable even at surface brightness levels as bright as 26.5 mag/arcsec$^2$.
   
LSB21. This galaxy has an elongated spheroidal shape. Its central surface brightness is  faint $\mu_g$(0)=26.12 mag/arcsec$^2$ and it has a rather flat radial colour profile \textsl{g}-\textsl{r}=0.4 mag. The surface brightness profiles of the galaxy show a change in the slope at around 16\arcsec, potentially indicating that this galaxy has two structural components with the latter resembling an exponential disc.
   
LBT1. This is the faintest galaxy in the sample discussed in the paper. It has a spheroidal morphology. It is invisible in the SDSS and Dragonfly datasets. Its central surface brightness is extremely low $\mu_g$(0)=27.11 mag/arcsec$^2$ and it has a rather flat radial colour profile \textsl{g}-\textsl{r}$\sim$0.65 mag. 

To conclude this section, we  expand further on the galaxy LBT1. This object is invisible in SDSS and Dragonfly, and therefore highlights the potential of spatial resolution, when combined with depth, for studying the low surface brightness universe. In this sense, the LIGHTS survey allows us to explore very faint and small galaxies around nearby (but well beyond the Local Group) galaxies such as NGC1042. In what follows, we  assume LBT1 is at the distance of NGC1042 (i.e. 13.5 Mpc or, equivalently, a distance modulus $\mu_0$=30.65 mag). However, for completeness, we also present its physical properties if the object is located at the distance of NGC1052 (i.e. 19 Mpc or $\mu_0$=31.39 mag). Nonetheless, we would like to point out the following. The distance between LBT1 and NGC1042 is 300\arcsec (i.e. 19.2 kpc at 13.5 Mpc) and 590\arcsec with respect to NGC1052 (i.e. 54.3 kpc at 19 Mpc). Considering the dynamical mass of NGC1042 (2.9$\times$10$^{10}$ M$_{\odot}$) and NGC1052 \citep[1.7$\times$10$^{12}$ M$_{\odot}$; ][]{2005MNRAS.358..419P}, it is possible to estimate the instantaneous tidal radius (r$_{tidal,inst}$) of LBT1 \citep[][]{2002AJ....124..127J}. To do such an estimation, it is necessary to assume a dynamical mass for LBT1. Here we use a range for the dynamical mass to light ratio between 10 to 100. This implies the following tidal radius for LBT1: 15\arcsec$<$r$_{tidal,inst}$$<$32\arcsec (caused by NGC1042) or 9\arcsec$<$r$_{tidal,inst}$$<$20\arcsec (caused by NGC1052). The absence of tidal distortions in LBT1 (at least in the inner 10\arcsec) favours, although not conclusively, the idea that this object is associated with NGC1042 rather than with NGC1052.

\begin{figure*}
   \centering
   \includegraphics[width=\textwidth]{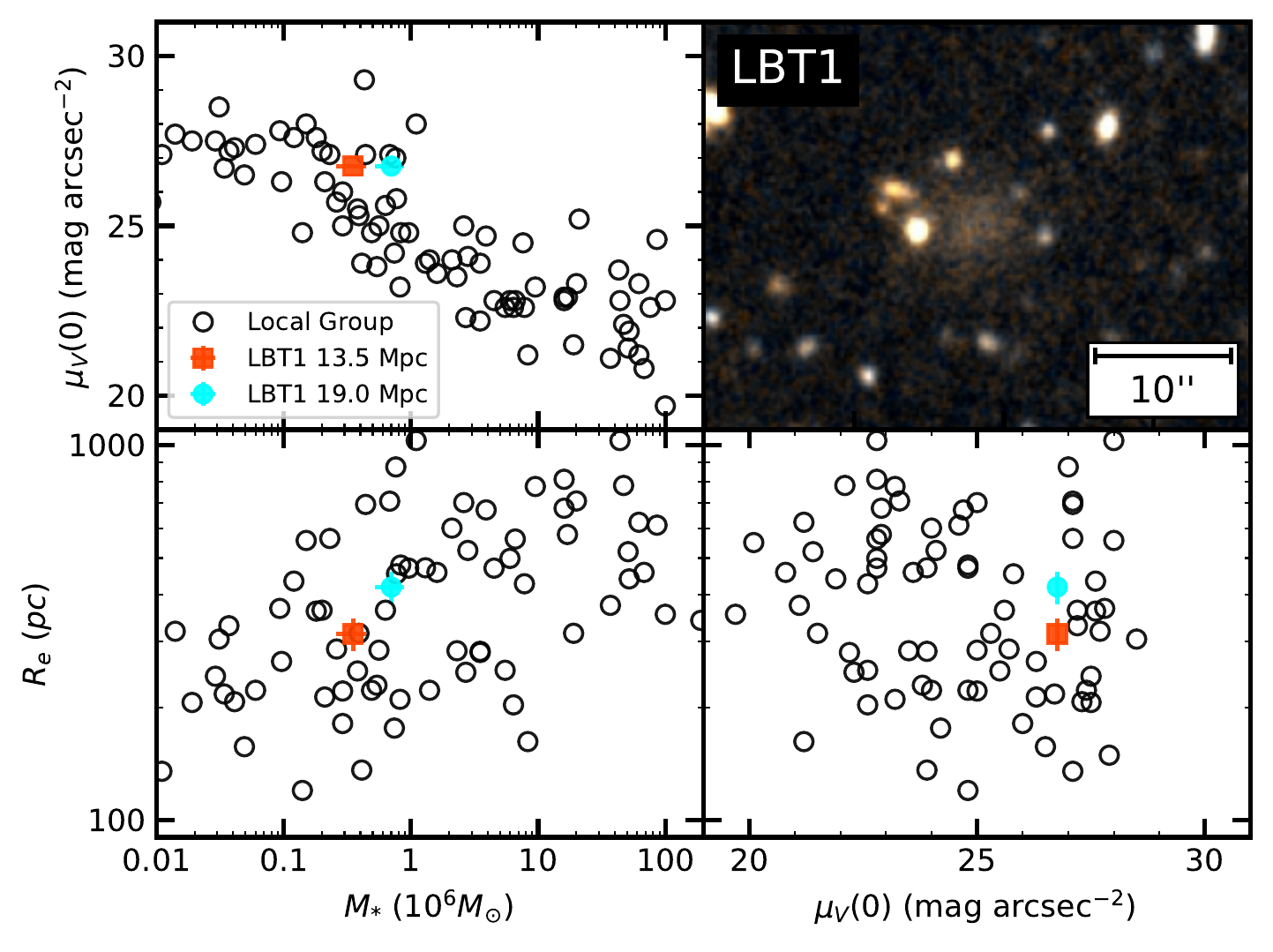}
      \caption{Structural properties of the galaxy LBT1 in relation to low mass galaxies in the Local Group \citep{2012AJ....144....4M}. For its stellar mass, LBT1 has an effective radius which is common within the Local Group satellite population. However, its central surface brightness ($\mu_V$(0)$\sim$26.8 mag/arcsec$^2$) is among the faintest for its stellar mass. If LBT1 were in the Local Group it would resemble properties similar to the dwarf spheroidals Andromeda XIV or Sextans.}
         \label{lbt1properties}
   \end{figure*}

If LBT1 is at the distance of NGC1042, then its absolute magnitudes are: M$_{g}$=-7.95$\pm$0.02  and M$_{r}$=-8.59$\pm$0.02 mag (M$_{g}$=-8.69$\pm$0.02  and M$_{r}$=-9.33$\pm$0.02 mag at 19 Mpc). Considering its global colour (\textsl{g}-\textsl{r}=0.64$\pm$0.03 mag), its (M/L)$_r$ would be 1.78$\pm$0.44$\Upsilon_{\odot}$ (assuming a Chabrier IMF) and therefore its total stellar mass is 3.5$\pm$0.9$\times$10$^5$ M$_{\odot}$ (7.0$\pm$1.8$\times$10$^5$ M$_{\odot}$ at 19 Mpc) and its effective radii are R$_{e,g}$=290$\pm$30 and R$_{e,r}$=338$\pm$30 pc (R$_{e,g}$=386$\pm$40 and R$_{e,r}$=451$\pm$40 pc at 19 Mpc). In Fig. \ref{lbt1properties}, we show the location of this galaxy in comparison to other dwarf galaxies in the Local Group \citep{2012AJ....144....4M}. To estimate both R$_e$ (in the \textsl{V}-band) and $\mu_{V}(0)$ for LBT1 we have interpolated between the values we have retrieved in the \textsl{g} and \textsl{r} bands. Examples of Local Group galaxies that are similar in stellar mass, effective radius and central surface brightness to LBT1 are the Andromeda XIV \citep{2007ApJ...670L...9M} dwarf spheroidal (if LBT1 is at 13.5 Mpc) or the Sextans \citep{1990MNRAS.244P..16I} dwarf spheroidal (if LBT1 is located at 19 Mpc). Both at 13.5 and 19 Mpc, LBT1 falls on top of the local scaling relationships. Therefore, we cannot favour either of these two distances with this photometric data alone.

As shown in Fig. \ref{lbt1properties}, LBT1 has an effective radius that is comparable to those satellites in the Local Group with similar stellar mass. However, the central surface brightness of LBT1 ($\mu_V$(0)$\sim$26.8 mag/arcsec$^2$) is among the faintest for its stellar mass. In fact, the object is around 10 times fainter than the average satellite galaxy with the same stellar mass in the Local Group. This highlights the power of LBT imaging for detecting even the faintest satellites beyond the Local Volume (D$>$12 Mpc).

\section{Discussion and conclusions}

In this paper we present the first results of the LIGHTS survey conducted with the LBT. This work shows the potential of LBT to conduct low surface brightness studies at extremely faint levels ($\mu_V$$\sim$31 mag/arcsec$^2$; 3$\sigma$ in 10\arcsec$\times$10\arcsec). The relatively large FOV of the LBC instrument (23\arcmin$\times$25\arcmin), the large collecting area of the LBT (2$\times$8.4m) and the possibility of conducting simultaneous observations in two different filters make LBT extremely competitive for these types of studies. Here we have shown that with only 1.5h on source, the telescope is capable of providing images that have comparable depths to those expected once the LSST survey is completed \citep[i.e. $\mu_g$$\sim$31.1 mag/arcsec$^2$ and $\mu_r$$\sim$30.6 mag/arcsec$^2$; 3$\sigma$ in 10\arcsec$\times$10\arcsec ][]{2018arXiv181204897L}. Not only the depth, but also the pixel scale of the future LSST and LBT images is  approximately the same (i.e. $\sim$0.2\arcsec). Together with the good quality seeing of both observatories, this allows exquisite spatial ground-based resolution. This superb spatial resolution is an enormous advantage for low surface brightness studies because the high spatial resolution allows one to detect, characterise and remove the many compact foreground (i.e. stars) and background (i.e. high-z galaxies) sources that plague such deep images. 

These similarities  allow us to preview what kind of low surface brightness studies the future LSST will be capable of. In this paper, we have focused on two important issues in extragalactic astronomy (namely the characterisation of  stellar haloes and of low mass satellites) that will  strongly benefit from having access to low surface brightness images. Before the completion of the LSST survey, the LIGHTS sample will constraint the stellar halo populations and satellite luminosity functions of dozens of nearby galaxies. In addition to this, we expect enormous progress is also possible on those other studies that require very deep imaging such as Zodiacal light \citep[see e.g.][]{2020P&SS..19004973L}, astrospheres \citep[see e.g.][]{2012A&A...537A..35C}, Galactic cirri \citep[see e.g.][]{2016A&A...593A...4M,2020A&A...644A..42R},  truncations and warps in galaxies \citep[see e.g.][]{2007A&A...466..883V,2012MNRAS.427.1102M,2016ApJ...826...59W}, and intra-cluster light \citep[see e.g.][]{2005ApJ...631L..41M,2005ApJ...618..195G,2014ApJ...794..137M,2018MNRAS.474..917M}. In this work, we have shown that current deep imaging with our largest ground-based telescopes  allows us to find and characterising the bi-dimensional structure of stellar haloes  even around modest mass (M33-like) galaxies well beyond the Local Volume. In addition, this dataset probe their population of low mass satellites (few times 10$^5$ M$_{\odot}$). This advance in deep imaging  promises to transform our understanding of galaxy formation.

\begin{acknowledgements}

We thank the referee for a constructive report that helped on improving the clarity of this manuscript. We acknowledge support from grant PID2019-107427GB-C32 from The Spanish Ministry of Science and Innovation. We acknowledge financial support from the European Union's Horizon 2020 research and innovation programme under Marie Sk\l odowska-Curie grant agreement No 721463 to the SUNDIAL ITN network, and the European Regional Development Fund (FEDER), from IAC project P/300624, financed by the Ministry of Science, Innovation and Universities, through the State Budget and by the Canary Islands Department of Economy, Knowledge and Employment, through the Regional Budget of the Autonomous Community. DZ acknowledges financial support from NSF AST-2006785. NC acknowledges support from the research project grant ‘Understanding the Dynamic Universe’ funded by the Knut and Alice Wallenberg Foundation under Dnr KAW 2018.0067 and Chris Usher for interesting comments. DJS acknowledges support from NSF grants AST-1821967 and 1813708. JR acknowledge funding from the State Agency for Research of the Spanish MCIU through the `Center of Excellence Severo Ochoa' award to the Instituto de Astrof{\'i}sica de Andaluc{\'i}a (SEV-2017-0709), financial support from the grants AYA2015-65973-C3-1-R and RTI2018-096228-B-C31 (MINECO/FEDER, UE) as well as support from the State Research Agency (AEI-MCINN) of the Spanish Ministry of Science and Innovation under the grant `The structure and evolution of galaxies and their central regions' with reference PID2019-105602GB-I00/10.13039/501100011033. The LBT is an international collaboration among institutions in the United States, Italy and Germany. LBT Corporation partners are: The University of Arizona on behalf of the Arizona Board of Regents; Istituto Nazionale di Astrofisica, Italy; LBT Beteiligungsgesellschaft, Germany, representing the Max-Planck Society, The Leibniz Institute for Astrophysics Potsdam, and Heidelberg University; The Ohio State University, representing OSU, University of Notre Dame, University of Minnesota and University of Virginia. This research has made use of the NASA/IPAC Extragalactic Database (NED), which is funded by the National Aeronautics and Space Administration and operated by the California Institute of Technology. This work was partly done using GNU Astronomy Utilities (Gnuastro, ascl.net/1801.009) version 0.13.12-f50c. Work on Gnuastro has been funded by the Japanese Ministry of Education, Culture, Sports, Science, and Technology (MEXT) scholarship and its Grant-in-Aid for Scientific Research (21244012, 24253003), the European Research Council (ERC) advanced grant 339659-MUSICOS, and the Spanish Ministry of Economy and Competitiveness (MINECO) under grant number AYA2016-76219-P.

\end{acknowledgements}


\bibliographystyle{aa}
\bibliography{lbt}

\end{document}